\documentclass[12pt]{article}
\usepackage[utf8]{inputenc}
\usepackage[english]{babel}
\usepackage{cmap}
\usepackage{amsmath}
\usepackage{graphicx} 
\usepackage{titlesec}
\usepackage{extsizes}
\usepackage{mathtools}
\usepackage{nicefrac}
\usepackage{upgreek}
\usepackage{authblk}
\usepackage{caption}
\usepackage{subcaption}

\title{Localized heat perturbation in harmonic 1D crystals. Solutions for an equation of anomalous heat conduction }
\author[1]{Aleksei A. Sokolov \thanks{sokolovalexey1@gmail.com} }
\author[1,2]{Anton M. Krivtsov}
\author[3]{ Wolfgang H. M\"uller}
\affil[1]{Peter the Great Saint Petersburg Polytechnic University}
\affil[2]{Institute for Problems in Mechanical Engineering RAS}
\affil[3]{Institute of Mechanics, Chair of Continuum Mechanics and Constitutive Theory, Technische
Universität Berlin}
\date {16 February 2017}

\titlespacing*{\section}      {0pt}{3.50ex plus 1ex minus .2ex}{2.3ex plus .2ex}
\titlespacing*{\subsection}   {0pt}{3.25ex plus 1ex minus .2ex}{1.5ex plus .2ex}
\titlespacing*{\subsubsection}{0pt}{3.25ex plus 1ex minus .2ex}{1.5ex plus .2ex}

\usepackage[top=2cm, bottom=2cm, left=3cm, right=1cm]{geometry}
\newcommand{\mr}{\mathrm}

\newcommand{\picscale}{0.7}
\DeclareMathOperator\erf{erf}

\begin{document}

\maketitle

\begin{abstract}
 In this work exact solutions for the equation that describes anomalous heat propagation in 1D harmonic lattices are obtained. Rectangular, triangular, and sawtooth initial perturbations of the temperature field are considered. The solution for an initially rectangular temperature profile is investigated in detail. It is shown that the decay of the solution near the wavefront is proportional to $1/ \sqrt{t}$. In the center of the perturbation zone the decay is proportional to $1/t$. Thus the solution decays slower near the  wavefront, leaving clearly visible peaks that can be detected experimentally.

\end{abstract}
\section{Introduction}
\par Nowadays investigation of nonlinear thermomechanical processes in low-dimensional structures attracts high interest due to the rapid development of nanoelectronical devices based on materials with microstructure \cite{6038223, Chen:2015aa, Goldstein2007235, PhysRevLett.93.184301}. Achievements in nanotechnology allowed for an experimental proof of the wave nature and finite propagation velocity of thermal perturbations \cite{doi:10.1063/1.1993768, Wang787}. This can provide a foundation for a universal theory of thermal conduction, applicable both on micro and macroscales.
\par The classical heat equation is a parabolic partial differential equation that describes the distribution of heat in a given region over time,
\begin{equation}
\label{eq:heat_eq}
	\dot{T} = \beta T'',
\end{equation}
where $T$ is temperature, $\beta$ is thermal diffusivity, dot $\dot{(~~)}$ denotes differentiation with respect to $t$, prime $(~~)'$ denotes differentiation with respect to $x$. The classical heat equation is derived on the basis of the Fourier's law \cite{cannon, lepri},
\begin{equation}
\label{eq:fourier}
	q = -\kappa \nabla T,
\end{equation}
where $\kappa$ is thermal conductivity, $q$ is heat flux, and $T$ is temperature. Practical applications show that at the macroscale Fourier's law of heat conduction is well applicable in order to describe temperature processes.  However, it predicts an infinite speed of signal propagation, which is paradoxical from a physical point of view. A study of processes at the microscale, when the characteristic length is proportional to several atomic bond lengths, requires us to use more complicated models of heat transfer, which take the finite velocity of heat propagation into account. Well observed abnormalities from the Fourier's law happen in thermal processes occurring in one-dimensional crystalline structures~\cite{hsiao}. Recent experimental works show length dependence of thermal conductivity of nanostructures~\cite{Zhang:2015aa}. Significant deviations from Fourier's law were shown for~\ensuremath{\mathrm{C}} and~\ensuremath{\mathrm{BN}} nanotubes ~\cite{PhysRevLett.101.075903}. Thermal anomalies for nanoscale structures can be used in practice for designing perspective devices such as thermal diodes \cite{PhysRevLett.93.184301}.  
\par The anomalous nature of heat processes for one-dimensional lattices was demonstrated analytically in paper~\cite{lebowitz}, 
where a problem of heat flow between two heat baths was considered. A hyperbolic heat equation is one of the alternatives to describe heat processes, which take finite speed of temperature propagation~\cite{cattaneo, vernotte} into account,
\begin{equation}
\label{eq:HCT}
	\tau \ddot{T} + \dot{T} = \beta T'',
\end{equation}
where $\tau$ is a relaxation time. However, Eqn. \eqref{eq:HCT} has serious difficulties in describing heat transfer in one-dimensional crystals, since no unique relaxation time can be determined \cite{gendelmann_FPU}.
A perspective approach for description of un-steady heat processes in 1D crystals is presented in papers \cite{  Krivtsov_DAN_2014, krivtsov, Krivtsov_arXiv}. By using correlational analysis, the initial stochastic problem for individual particles is reduced to a deterministic problem for statistical characteristics of the crystal motion. Finally, a continuum equation \eqref{eq:temperature_equation} describing anomalous heat transfer in 1D harmonic lattices is obtained in paper \cite{krivtsov}. In the current work exact solutions for this equation will be obtained. Solutions for a number of problems such as rectangular, triangular, and sawtooth initial perturbations will be obtained. Properties of the solution for rectangular initial perturbation such as decay behavior and asymptotics of the wavefront will be investigated. These results can be used for analysis of the anomalous heat transfer in more complex systems, such as 1D crystals on elastic foundation \cite{Babenkov} and 2D-3D crystals \cite{Kuzkin_Krivtsov}. An understanding of the anomalous heat conduction is important for analysis of the experimental results, which are to be obtained in the nearest future due to the rapid development of nanotechnologies. 
\section{Localized perturbations in a harmonic chain}
\par The harmonic chain is a simple and powerful model in order to investigate anomalous heat conduction phenomena. Following on to the work \cite{krivtsov} let us consider an infinite harmonic chain. Each particle with mass $m$ is connected to its neighbor by Hookean springs with stiffness $C$. The equation of motion of the particles reads:
\begin{equation}
	\ddot{u}_k = \omega_e^2( u_{k-1} - 2 u_k + u_{k+1}), \quad \omega_e \stackrel{\mathclap{\normalfont\mbox{def}}}{=} \sqrt{ \frac{C}{m}},
\end{equation}
where $u_k$ is displacements of particle with index $k$. The following initial conditions are considered:
\begin{equation}
\label{eq:particles_initial}
	u_k|_{t = 0} = 0, \qquad \dot{u}|_{t=0} = \sigma(x) \rho_k,
\end{equation}
where $\rho_k$ are independent random variables with zero expectation and unit variance; $\sigma$ is the variance of the initial particle velocity. The variance is a slowly changing function of the spatial coordinate $x = ka$, where $a$ is the initial distance between neighboring particles. Such initial conditions can be realized by ultrafast heating, for example with a laser \cite{experimental}. Let us introduce  the kinetic temperature $T$ as
\begin{equation}
	k_BT = m  \langle \dot{u_k} \rangle^2,
\end{equation}
where $\langle ... \rangle$ is an operator averaging over realizations, and $k_B$ is the Boltzmann constant.
In paper~\cite{krivtsov} a continuum partial differential equation for the kinetic temperature was obtained:
\begin{equation}
\label{eq:temperature_equation}
	\ddot{T} + \frac{1}{t} \dot{T} = c^2 T'',
\end{equation}
where $c$ is the speed of sound in a one-dimensional crystal. Eqn.~\eqref{eq:temperature_equation} describes the evolution of the spatial temperature distribution in the chain. The following initial conditions for the equation~\cite{krivtsov} corresponds to stochastic initial conditions~\eqref{eq:particles_initial}:
\begin{equation}
\label{eq:initial}
	\dot{T}|_{t=0} = 0, \qquad T|_{t=0} = T_0(x).
\end{equation}
The solution of the initial problem  \eqref{eq:temperature_equation}--\eqref{eq:initial} can be obtained in  integral form \cite{krivtsov}:
\begin{equation}
\label{eq:temperature}
	T(t,x) = \frac{1}{\pi} \int \limits_{-t}^t \frac{T_0(x - c\tau)}{\sqrt{ t^2 - \tau^2}}\mr{d}\tau.
\end{equation}
Eqn. \eqref{eq:temperature_equation}  is a particular case of the Darboux equation~\cite{darboux}. This type of equation was investigated earlier in context with spherical averages for solutions of 2D and 3D wave equations. However, it was not investigated well in connection with the problems of heat conduction.  Eqn. \eqref{eq:temperature_equation} looks similar to the hyperbolic heat equation \eqref{eq:HCT}, however, it has a variable coefficient. This peculiarity is due to anomalous heat transfer in a 1D chain. From the form of Eqn. \eqref{eq:temperature_equation} it seems that it has a singularity. However, it does not matter because Eqn. \eqref{eq:temperature_equation} is to be solved together with the initial conditions \eqref{eq:initial}. The absence of singularity is confirmed by the general analytical solution \eqref{eq:temperature} and solutions of particular problems that will be considered below.
\par This work is dedicated to finding exact analytical solutions of Eqn.~\eqref{eq:temperature_equation} for cases when the initial thermal distribution $T_0(x)$ is a localized function of coordinate~$x$,
\begin{equation}
\label{eq:arbitrary_initial}
	T_0(x) = \begin{cases}
			0, &x < -l, \\
			\varPhi(x), &-l< x <l, \\
			0, &x > l,
		\end{cases}
\end{equation}
where $\varPhi(x)$ is an arbitrary function, and $l$ is the half width of the localized perturbation.
Experimentally such an initial temperature distribution \eqref{eq:arbitrary_initial}  can be realized by performing superfast laser heating of a localized region of the chain.
\section{Rectangular perturbation}
\subsection{Solution}
Let us consider the case when the initial temperature perturbation has a rectangular shape:
\begin{equation}
\label{eq:initial_step}
	T_0(x) = A\left( \mathcal{H}\left(x+l \right) - \mathcal{H}\left(x-l\right) \right),
\end{equation}
where $\mathcal{H}(x)$ is the Heaviside function:
\begin{equation}
	\mathcal{H}(x) = \begin{cases}
			0, &x < 0, \\
			1, &x \ge 0.
		\end{cases}
\end{equation}
$A$ is the amplitude of the temperature perturbation. After substituting formula \eqref{eq:initial_step} into the solution \eqref{eq:temperature} we obtain:
\begin{equation}
\label{eq:sum_heavy}
	T(t,x) = 
  \frac{A}{\pi} \int \limits_{-t}^t \frac{ \mathcal{H}(x+l)}{\sqrt{ t^2 - \tau^2}}\mr{d}\tau  -  \frac{A}{\pi} \int \limits_{-t}^t \frac{ \mathcal{H}(x-l)}{\sqrt{ t^2 - \tau^2}}\mr{d}\tau.
\end{equation}
By substituting the solution for a single Heaviside initial impulse, which was obtained in \cite{krivtsov},
\begin{equation}
 T(x, t) \, \stackrel{\mathclap{\normalfont\mbox{def}}}{=} \,  T_S(x, t)  =\begin{cases}
						
						0, & x \le - ct,\\
						\cfrac{A}{\pi} \arccos\left(\cfrac{x}{ct}\right)& - ct \le x \le ct,\\
						A, &   x \ge ct
					
					\end{cases}
\end{equation}
to  \eqref{eq:sum_heavy} we obtain the solution of the given problem as a linear combination of these solutions. Solution for positive $x$:

\begin{equation}
\label{eq:step_solution0}
t \le \tau_0: \qquad T(x, t) = \begin{cases}
                                                      0, &  l + ct \le x,\\
                                                      \frac{A}{\pi} \arccos(\frac{x-l}{ct}), &l - ct \le x \le l + ct,\\
						A, & 0 \le x \le l -ct,\\
						
					\end{cases}\hspace{3cm}
\end{equation}

\begin{equation}
\label{eq:step_solution}
t \ge \tau_0: \qquad T(x, t) = \begin{cases}
						0, &ct+l \le x,\\
						\frac{A}{\pi} \arccos(\frac{x-l}{ct}),& ct - l \le x \le ct + l,\\
						\frac{A}{\pi} \left( -\arccos(\frac{x+l}{ct}) + \arccos\frac{x-l}{ct} \right), & 0 \le x \le ct -l,\\
					\end{cases}
\end{equation}
where $\tau_0 = l/c$. For negative $x$ the solution is symmetric and can be obtained by $T(x, t) = T(-x,t)$. \par
\begin{figure*}
\centering
\begin{subfigure}{.48\textwidth}
  \centering
  \includegraphics[width=1 \linewidth]{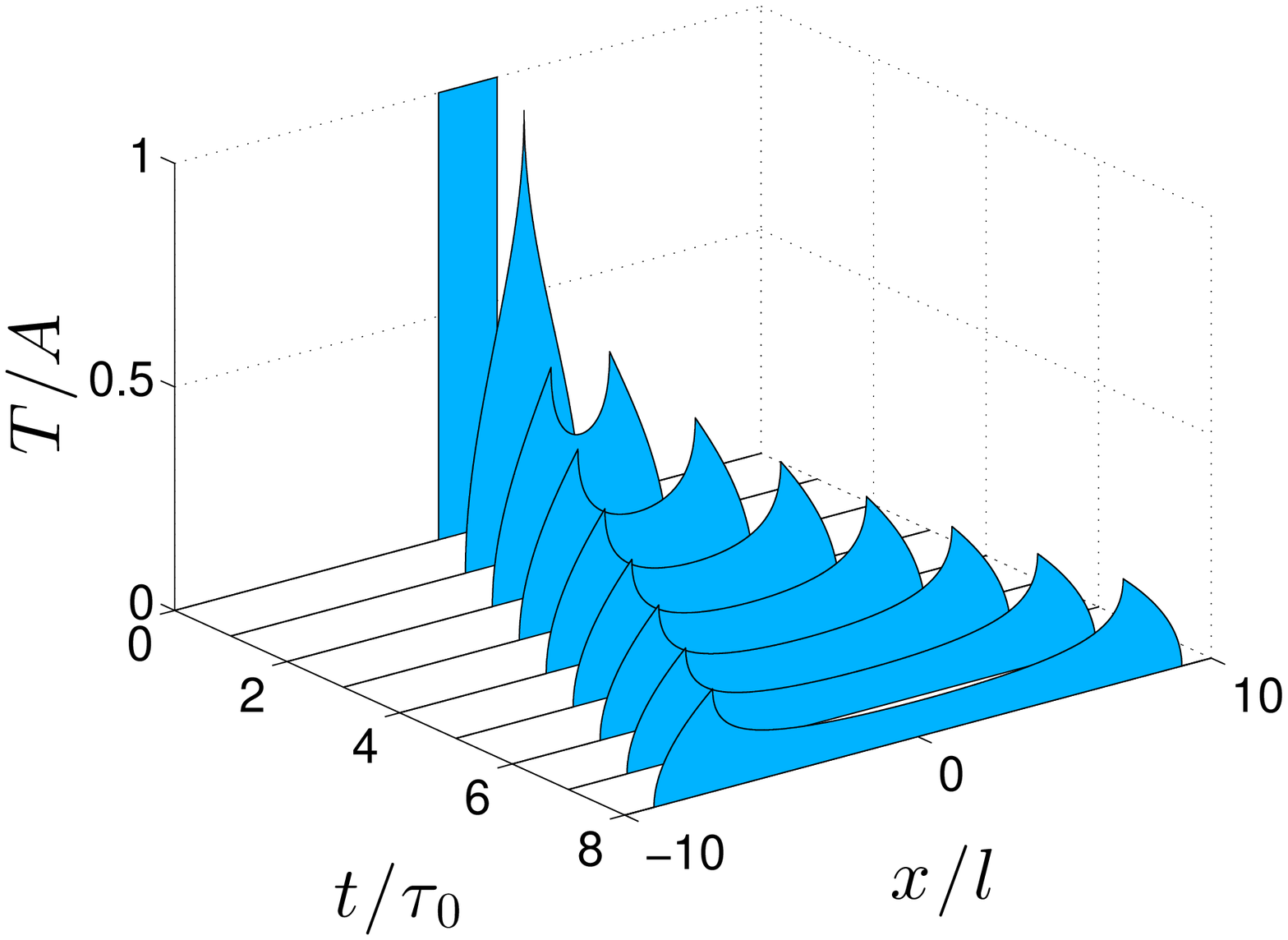}
  \caption{Anomalous}
  \label{fig:sub1}
\end{subfigure}
\begin{subfigure}{.48\textwidth}
  \centering
  \includegraphics[width=1 \linewidth]{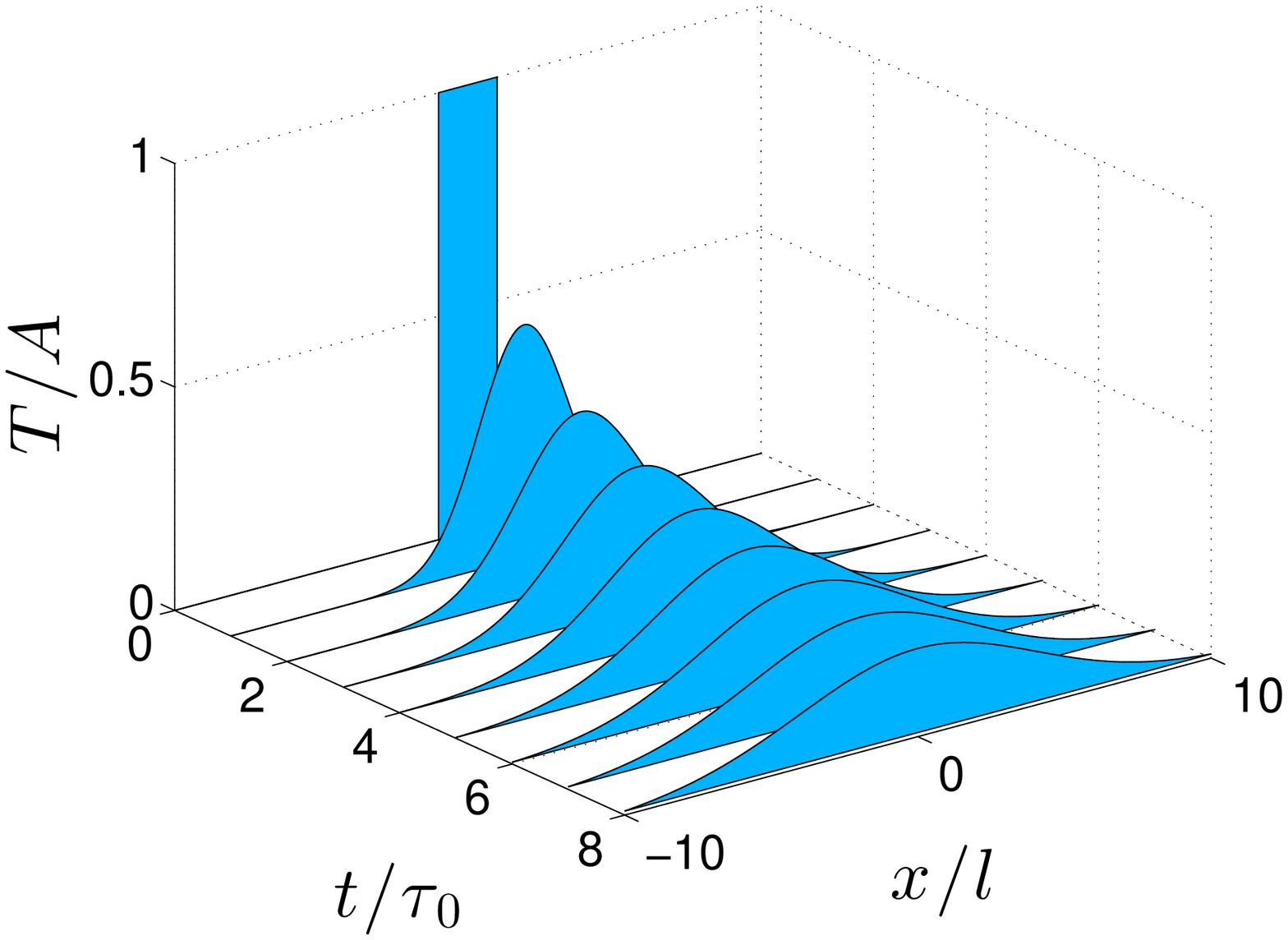}
  \caption{Fourier}
  \label{fig:sub2}
\end{subfigure}%
\caption{Time evolution for solutions for a rectangular initial perturbation.}
\label{pic:step_solution}
\end{figure*}

 For comparison let us consider the same initial problem for the classical heat equation:
\begin{equation}
\label{eq:fourier}
	\dot{T} = \beta T''.
\end{equation}
The solution for an initial Heaviside step temperature perturbation has the form~\cite{Mueller_Fundamentals}
\begin{equation}
\label{eq:fourier_step_solution}
	T(x,t) = \frac{1}{2}\erf \left( \frac{x}{ \sqrt{ 4 \beta t}} \right),
\end{equation}
where $\erf(x)$ is the Gaussian error function. Then solution of the initial problem \eqref{eq:initial}, \eqref{eq:initial_step}, \eqref{eq:fourier} is:
\begin{equation}
	T(x,t) = \frac{1}{2}\erf \left( \frac{x+l}{ \sqrt{ 4 \beta t}} \right) - \frac{1}{2} \erf \left( \frac{x-l}{ \sqrt{ 4 \beta t}} \right).
\end{equation}
	The time evolution plots for the solution of anomalous heat equation~\eqref{eq:temperature_equation} and the Fourier equation~\eqref{eq:fourier} are shown in Fig. \ref{pic:step_solution}. Let us compare the two solutions. The Fourier solution is forming a peak at $x=0$ which decays exponentially. For the case of the anomalous heat equation the solution decays in the area near $x=0$ more rapidly than near the wavefronts forming two peaks. The peaks travel in negative and positive directions with coordinates $x = -l + ct$ and $x = l - ct$. 

\subsection{Decay behavior}
Let us consider the decay behavior of the solution \eqref{eq:step_solution} at  $x = 0$. We perform a series expansion of the solution,
\begin{equation}
	T(t, 0) = \frac{A}{\pi} \left[ \pi - 2\arccos\left( \frac{l}{ct} \right) \right] =2\varepsilon + O( \varepsilon^3),
\end{equation}
where $\varepsilon = \frac{l}{ct}$ is a small parameter.

Now let us consider the decay behavior of the peaks $ x = l -ct$ and $ x = -l + ct$. From formula~\eqref{eq:step_solution} it follows that:

\begin{equation}
\label{eq:max_fading_law}
	T(t, -l + ct) = T(t, l -ct)= \frac{A}{\pi}\left[ \pi - \arccos\left( \frac{2l}{ct} - 1\right) \right] = 2 \sqrt{ \varepsilon } +    O\left( \varepsilon^\frac{3}{2} \right),
\end{equation}

Summarizing the above:
\begin{equation}
	T(t, 0) \stackrel { t \rightarrow \infty} {\sim} 2\varepsilon \sim \frac{1}{t}\;, \qquad T(t, -l + ct) = T(t, l - ct) \stackrel{t \rightarrow \infty}{ \sim } 2 \sqrt{\varepsilon}  \sim \frac{1}{ \sqrt{t}}.
\end{equation}
Thus, the solution decays faster in the area between wavefronts (proportional to $1/t$) rather than near the wavefront (proportional to $1/ \sqrt{t}$). Thus the peaks remain strongly pronounced even for long times.

\subsection{Envelope curve for the peaks}
The solution \eqref{eq:step_solution} has two peaks. The peaks travel in positive and negative directions at speed $c$. Since the solution is symmetric let us consider only the peak with the coordinate $x = ct-l$.  We shall consider the curve drawn by the peak of the solution as it travels in positive direction. By substituting $t = \frac{x + l}{c}$ into formula \eqref{eq:max_fading_law}  we obtain the expression for the enveloping curve:
\begin{equation}
	T_{\mr{env}}(x)= \frac{A}{\pi} \left[ \pi - \arccos \left( \frac{2l}{x+l} - 1\right) \right].
\end{equation}
For any $x$ we have: $T(x) \le T_{\mr{env}} \left( |x| \right)$.
\begin{figure*}[h]
\center{\includegraphics[width= 1 \linewidth]{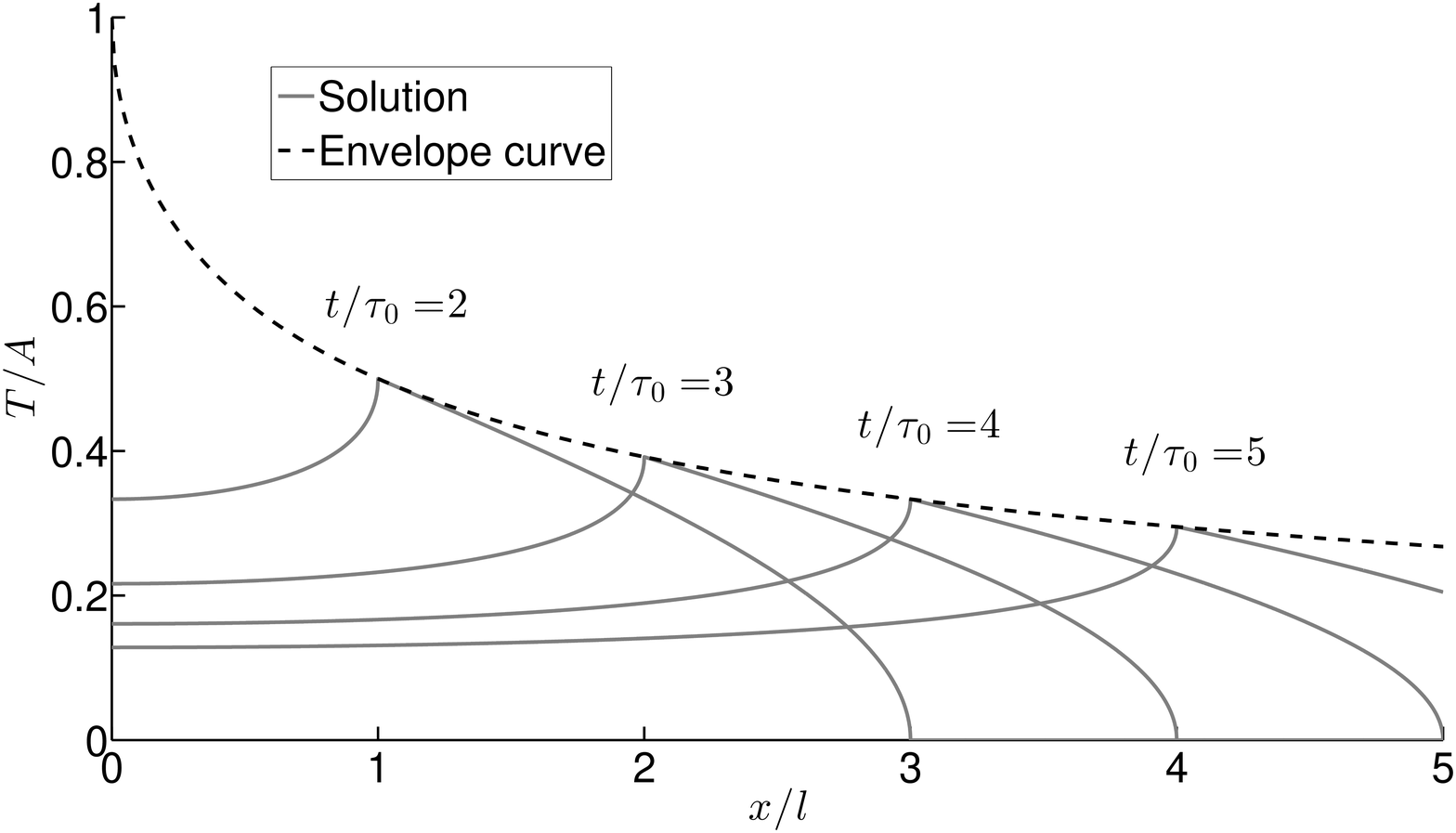}}
\caption{ Enveloping curve for peaks of solution.}
\label{pic:envelope}
\end{figure*}
The enveloping curve is shown in Fig.\ref{pic:envelope}. The expression decays as $1/\sqrt{x}$, which corresponds to the fact that the solution decays as $1/ \sqrt{t}$ near the wavefront (the wavefront travels at constant speed).

\subsection{Asymptotic behavior of the wavefront}
\begin{figure*}[h]

\center{\includegraphics[width= 1 \linewidth]{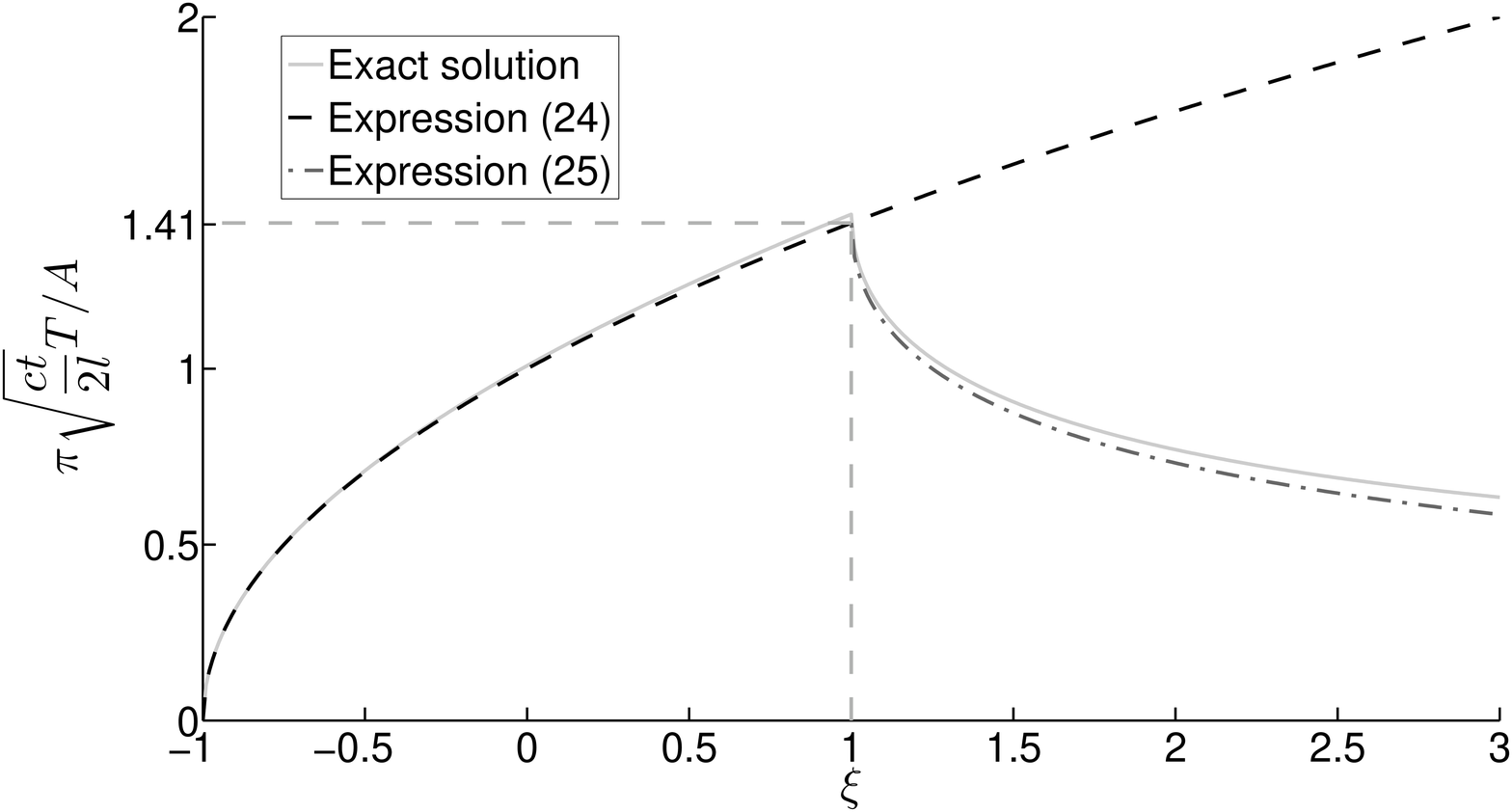}}
\caption{ Approximation curves for rectangular initial temperature perturbation.}
\label{pic:step_wavefront_approx}
\end{figure*}

Let us consider the solution \eqref{eq:step_solution} near the wavefront at long times $t$.  Let $\xi = \frac{-x + ct}{l}$. For $ x \in [-l+ct; l+ct] $ we have:
\begin{equation}
\label{eq:rect_front_approx1}
	T(\xi, t) = \frac{A}{\pi} \arccos \left[ 1 - \frac{l}{ct}( \xi + 1) \right]  \stackrel{ t \rightarrow \infty } {\sim}  \frac{A}{\pi} \sqrt{ \frac{2l}{ct}} \sqrt{ \xi + 1},
\end{equation}
where $\cfrac{\xi + 1}{ct}$ is a small parameter, used for expansion.
For $ x \in [l-ct; -l+ct]$ we have:
\begin{equation}
\label{eq:rect_front_approx2}
	\begin{aligned}
	T(\xi, t) = &\frac{A}{\pi} \left( - \arccos \left[ 1 - \frac{l}{ct} (\xi -1) \right] + \arccos \left[ 1- \frac{l}{ct}(\xi+1)  \right] \right)  \stackrel{ t \rightarrow \infty } {\sim} \\  \stackrel{ t \rightarrow \infty } {\sim} & \frac{A}{\pi} \sqrt{ \frac{2l}{ct}}\left( \sqrt{\xi+1} - \sqrt{\xi-1} \right).
	\end{aligned}
\end{equation}
The functions \eqref{eq:rect_front_approx1} and \eqref{eq:rect_front_approx2} have the following structure:
\begin{equation}
\label{eq:step_assympt}
T = \cfrac{A}{\pi}\sqrt{ \cfrac{2l}{ct} }\,F(\xi).
\end{equation}
The plot of the solution \eqref{eq:step_solution}, expressions  \eqref{eq:rect_front_approx1} and  \eqref{eq:rect_front_approx2} with the corresponding dimensionless time parameter $t/\tau_0 = 100$ is shown in Fig.~\ref{pic:step_wavefront_approx}. The relation \eqref{eq:step_assympt} means that the shape of the solution shrinks vertically with time, but it does not change horizontally. The asymptotic solutions \eqref{eq:rect_front_approx1} and \eqref{eq:rect_front_approx2} give peak values of $T$ for $\xi=1$, where $F(\xi) = \sqrt{2}$. Thus the solution at this point is continuous, but not smooth (the derivative of the solution has a jump).

\section{Triangular perturbation}
In order to obtain the solution for a triangular initial function we consider the following auxiliary problem where the initial temperature distribution is a linearly heated semispace,
\begin{equation}
\label{eq:initial_linear}
	T_0(x) = \begin{cases}
	0 , &x < 0 ;\\
	 Bx, &x \ge 0,
		 \end{cases}
\end{equation}
where $B = A/l$ is a constant of proportionality.
After substituting \eqref{eq:initial_linear} in \eqref{eq:temperature} we obtain the solution for $ |x| < ct$,
\begin{equation}
\label{eq:semispace_solution}
	T(x,t) = Bx\left( \frac{1}{\pi} \arcsin x + \frac{1}{2} \right) + \frac{B}{\pi} \sqrt{t^2 c^2 - x^2} \quad \stackrel{\mathclap{\normalfont\mbox{def}}}{=} \quad f(x),
\end{equation}
and for $|x| > ct$ the initial temperature distribution is preserved.

\par
\begin{figure*}[h]
\center{\includegraphics[width= \picscale \linewidth]{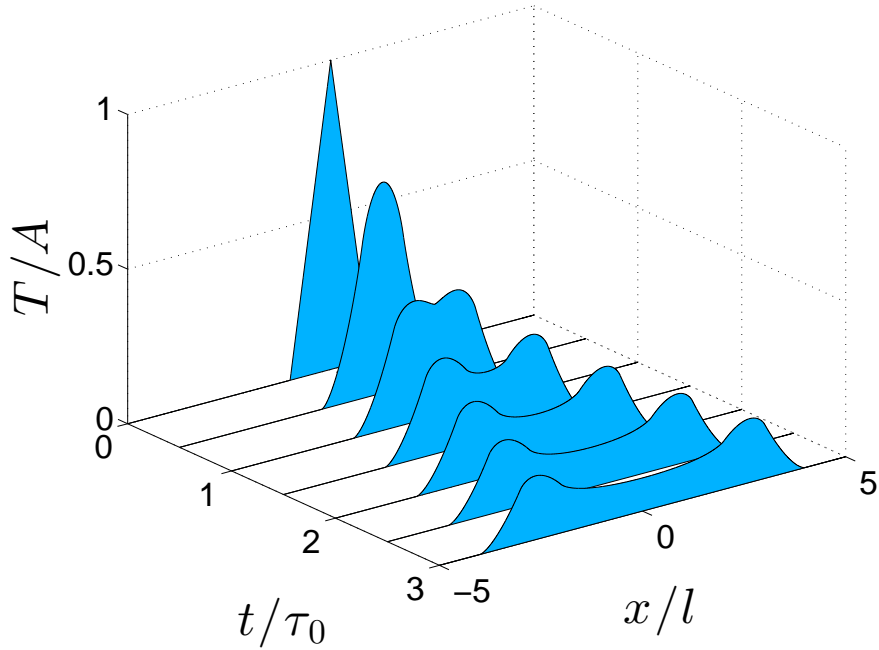}}
\caption{Time evolution of solution for a triangular initial perturbation.}
\label{pic:triangle_solution}
\end{figure*}

 Now we consider the problem for a triangular initial heat perturbation, which can be expressed by the following piecewise function:
\begin{equation}
\label{eq:triangle_initial}
	T_0(x) = \begin{cases}
			0, &x< -l ,\\
			(x+l)B, &-l \le x < 0,\\
			(-x+l)B, &0 \le x <l,\\
			0, &x \ge l.\\
		\end{cases}
\end{equation}
The solution for the initial temperature distribution \eqref{eq:triangle_initial} will be a linear combination of solutions for the linearly heated semispace. Denote the solution \eqref{eq:semispace_solution} by $T_L$. The solution for the initial distribution \eqref{eq:triangle_initial} will then be as follows:
\begin{equation}
\label{eq:triangle_solution}
	T(t,x) = T_L(t,x - l) + T_L(t, x+l)  - 2T_L(t, x),
\end{equation}
the solution is symmetric, $T(x,t) = T(-x,t)$. The part corresponding to positive $x$ has the following piecewise form:

\begin{align}
	t \le \tau_0/2 :\quad T(t,x) = \begin{cases}
								f(x+l) - 2f(x),& ~~~ 0 \le  x \le ct,\\
								(-x+l)B,&~~~ ct \le x \le l - ct,\\		
								(-x+l)B  + f(x+l),& ~~~ l - ct \le x \le l + ct,  \\
								0, &~~~  l + ct \le x,
					\end{cases}
\end{align}

\begin{align}
	\tau_0/2 \le t  \le \tau_0: \quad T(t,x) = \begin{cases}
								(x+l)B - 2f(x), & 0 \le x \le l  - ct, \\
								(x+l)B + f(x-l) - 2f(x),& l-ct \le x \le ct,\\		
								(-x+l)B+ f(x-l),& ct \le x \le l + ct, \\
								0,& l + ct \le x,
					\end{cases}
\end{align}

\begin{align}
	t \ge \tau_0:  \quad T(t,x) = \begin{cases}
								f(x+l) + f(x-l) - 2f(x), & 0 \le x \le -l + ct,\\
								(x+l)B + f(x-l) - 2f(x),& - l + ct \le x \le ct,\\		
								(-x+l)B + f(x-l), &  ct \le x \le l + ct,  \\
								0, & l + ct \le x.
					\end{cases}
\end{align}
The plot of the solution for the triangular initial perturbation is shown in Fig. \ref{pic:triangle_solution}.

Unlike the solution for a rectangular initial perturbation, which has a wavefront with vertical tangent and infinite derivative and a break of the temperature profile at the peak, the solution for a triangular perturbation has a smooth beginning at the wavefront and smooth behavior at the peak.

\section{Sawtooth perturbation}
\begin{figure*}[h]
\center{\includegraphics[width=\picscale  \linewidth]{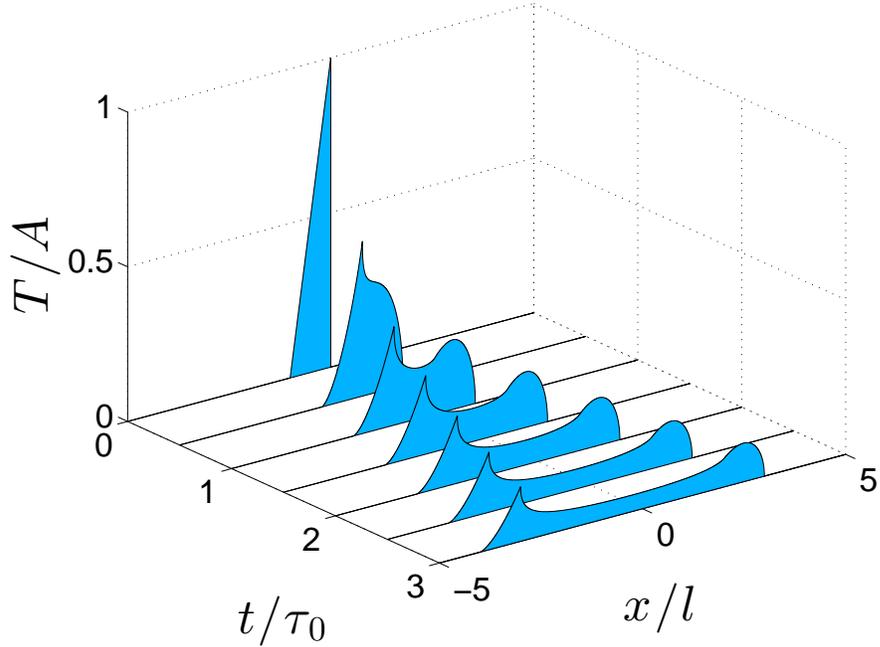}}
\caption{ Time evolution of solution for sawtooth initial perturbation.}
\label{pic:sawtooth_solution}
\end{figure*}

We consider an initial heat perturbation as a sawtooth spatial temperature distribution. It can be written in the following form:
\begin{equation}
\label{eq:sawtooth_initial}
\begin{aligned}
T_0(x) = \begin{cases}
		0, &\quad x \le -l ,\\
		x+l, &\quad -l \le x < 0,\\
		0, &\quad 0 \le x.\\
	\end{cases}
\end{aligned}
\end{equation}
The initial conditions \eqref{eq:sawtooth_initial} can be written as linear combinations of step function and linearly heated semispace. Then the solution for a sawtooth initial perturbation~\eqref{eq:sawtooth_initial} can be obtained from the corresponding combination of the solutions for a step initial distribution $T_S(x,t)$ and a linearly heated semispace initial distribution $T_L(x,t)$:
\begin{equation}
	T(t,x) = T_L(x+l,t) + T_L(x,t)+T_S(x,t),
\end{equation}
it has the following piecewise form:
\begin{equation}
t \le \tau_0: \quad T(x, t)=
					\begin{cases}			
						0, & x \le - l - ct,\\
						f (x+l), & -l - ct \le x \le -l + ct,\\
						B x, &-l + ct \le x \le  -ct,\\
						Bx - f(x) - \frac{A}{\pi} \arccos(\frac{x}{ct}), & -ct \le x \le ct,\\
						0, &x > ct,
					
					\end{cases}
\end{equation}

\begin{equation}
\label{eq:saw_solution}
	t \ge \tau_0: \quad T(x, t) = \begin{cases}
						
						0, &x \le -ct - l ,\\
						f (x+l)&  - ct -l \le x \le -ct,\\
						f (x+l) - f(x) - \frac{A}{\pi} \arccos(\frac{x}{ct}), & - ct \le x \le ct - l ,\\
						B x - f(x) - \frac{A}{\pi} \arccos(\frac{x}{ct}),& ct -l \le x \le ct,\\
						0, & ct \le x.
					
					\end{cases}
\end{equation}
The plot of the solution is shown in Fig.~\ref{pic:sawtooth_solution}. The left wavefront has a smooth beginning and an infinite derivative at the peak. On the other hand, the right wavefront has an infinite derivative and vertical tangent at the beginning, smooth behavior at the peak, and a horizontal tangent and zero derivative at the peak. 

\section{Conclusions}
The process of heat transfer in a 1D infinite harmonic chain was investigated. Localized initial perturbations were considered. Solutions for an equation describing anomalous heat conduction \eqref{eq:temperature_equation} derived in \cite{krivtsov} were obtained. Exact analytical solutions for rectangular, triangular, and sawtooth initial impulses were considered. It was shown that solutions for \eqref{eq:temperature_equation} unlike solutions for classical heat equation have a strongly pronounced wavefront. For the rectangular case it was shown that the decay of the solution near the wavefront is proportional to $1/\sqrt{t}$. Near zero the decay is proportional to $1/t$. Thus the solution decays slower near the wavefront, leaving clearly observable peaks. The shape of the wavefront is described by a function inversely proportional to the square root of time and has the form $T = \cfrac{A}{\pi}\,\sqrt{ \cfrac{2l}{ct} }\,F(\xi) $. \par The solution for a triangular initial temperature perturbations has a smooth beginning at the wavefront and a smooth behavior at the peak. \par In case of a sawtooth initial perturbation we have a non-symmetrical solution. The left wavefront has a smooth beginning and an infinite derivative and vertical tangent at the peak; the right wavefront has an infinite derivative and vertical tangent at the beginning, smooth behavior at the peak, zero derivative and a horizontal tangent at the peak.  \par The obtained solutions demonstrate the wave behavior accompanied with power decay. This differs from the results obtained from the solutions of the classic heat equation~\eqref{eq:heat_eq}~(diffusive behavior, exponential decay) and the hyperbolic heat equation~\eqref{eq:HCT}~(wave behavior, exponential decay). Such properties of the obtained solutions can be applied for analysis of the experimental data and choosing the right model for the description of the heat processes.
\section*{Acknowledgement}
The work was supported by Russian Science Foundation [Grant No. 14-11-00599].

\bibliography{sokolov_23_03_2017__arXiv_bib}

\end{document}